\documentclass[a4paper,12pt]{article}
\usepackage{amsmath}
\usepackage{amsfonts}
\usepackage{amssymb}
\usepackage{latexsym}
\usepackage{epsfig}
\usepackage{graphicx}
\usepackage{oldgerm}
\usepackage{theorem}
\usepackage{bm}

\def\R{\mathbb{R}}
\def\N{\mathbb{N}}

\def\r{\bm{r}}
\def\v{\bm{v}}
\def\F{\bm{F}}
\def\W{\bm{W}}
\def\w{\bm{w}}

\def\1{{\bf 1}}

\def\bra{\langle}
\def\ket{\rangle}

\theorembodyfont{\itshape}

\newtheorem{thm}{Theorem}[section]

\newtheorem{prop}[thm]{Proposition}

\newcommand{\SSC}[1]{\section{#1}\setcounter{equation}{0}}
\newcommand{\qed}{\hbox{\rule[-2pt]{3pt}{6pt}}}

\newfont{\bg}{cmr10 scaled\magstep4}

\newcommand{\bigzerou}{\smash{\lower1.7ex\hbox{\bg 0}}}

\begin{document}

\title{\Large{\bf
Interacting Particle Systems \\
Modeling Self-Propelled Motions
}
}
\author{
Saori Morimoto
\footnote{
Department of Physics,
Faculty of Science and Engineering,
Chuo University, 
Kasuga, Bunkyo-ku, Tokyo 112-8551, Japan
}, \, 
Makoto Katori
\footnote{
Department of Physics,
Faculty of Science and Engineering,
Chuo University, 
Kasuga, Bunkyo-ku, Tokyo 112-8551, Japan;
e-mail: 
makoto.katori.mathphys@gmail.com
}, \,
Hiraku Nishimori
\footnote{
Meiji Institute for Advanced Study of Mathematical Sciences, 
Meiji University, 
Nakano, Nakano-ku, Tokyo 164-8525, Japan;
e-mail: nishimor2@meiji.ac.jp
}
}

\date{13 February 2025}
\pagestyle{plain}
\maketitle

\begin{abstract}
In non-equilibrium statistical physics, active matters in both 
living and non-living systems have been extensively studied.
In particular, self-propelled particle systems provide challenging
research subjects in experimental and theoretical physics,
since individual and collective behaviors of units
performing persistent motions can not be described by
usual fluctuation theory for equilibrium systems.
A typical example of man-made self-propelled systems
which can be easily handled in small-sized experiments
is a system of camphor floats put on the surface of water.
Based on the experimental and theoretical studied
by Nishimori et al.~(J. Phys. Soc. Jpn. \textbf{86} (2017) 101012), 
we propose a new type of
mathematical models for complex motions of camphor disks
on the surface of water. 
In the previous mathematical models introduced by
Nishimori et al. are coupled systems of
the equations of motion for camphor disks described by 
ordinary differential equations
and the partial differential equation 
for the concentration field of camphor molecules in water.
Here we consider coupled systems of
equation of motions of camphor disks and
random walks representing individual 
camphor molecules in water.
In other words, we take into account non-equilibrium fluctuations
by introducing stochastic processes into the 
deterministic models.
Numerical simulation shows that our models can represent
self-propelled motions of individual camphor disk
as well as repulsive interactions among them.
We focus on the one-dimensional models 
in which viscosity is dominant, and 
derive a dynamical system of a camphor disk
by taking the average of random variables
of our stochastic system.
By studying both of stochastic models and
dynamical systems, we clarify the 
transitions between three phases of motions
for a camphor disk depending on parameters.

\vskip 0.2cm



\end{abstract}
\vspace{3mm}
\normalsize

\SSC
{Introduction}
\label{sec:introduction}

Although persistent motion is a common feature of
living systems, recently several models composing of 
self-propelled non-living units have been studied
to simulate physical and chemical systems.
Such \textit{self-propelled particle systems}
exhibit many interesting collective motions,
which can not be realized as the thermal or
chemical fluctuation phenomena in equilibrium systems.
Self-propelled particle systems
in both living and non-living worlds
have been extensively studied
in non-equilibrium statistical physics \cite{VZ12}.

In order to clarify the fundamental mechanism of individual 
and cooperative motions of self-propelled particles,
artificial systems consisting of easily handled particles
shall be studied experimentally and theoretically.
As a typical example among various types of
man-made self-propelled particles, a system of 
\textit{camphor floats}
put on the surface of water has been widely studied.
In particular, a camphor disk is the simplest
object which is a circular camphor float
\cite{MINS19,NSN17,STNN15}.
A camphor disk is prepared by soaking a filter paper
in camphor solution and drying it.
From a camphor disk floated on water,
camphor molecules are extended to the water surface,
which decreases the surface tension around the disk.

\vskip 0.3cm
\noindent
\textbf{Experimental Observation by Nishimori et al.~\cite{NSN17}}
\vskip 0.2cm

Nishimori et al.~\cite{NSN17} reported the following observation
in experiments, where a camphor disk was
floated in a quasi-one-dimensional water channel 
starting from the rest state.
The following are cited from \cite{NSN17}.

\begin{description}
\item[Initial condition]
At the initial condition,
no camphor molecule is present  on a water and camphor disk
is set in a stationary state (rest state).
Therefore, camphor concentration around the disk increases
with time due to the supply of camphor molecules from the disk,
which spread out on water and sublimate from the water surface to the air.
These three phenomena, i.e., supply, diffusion, and sublimation,
make a surface-concentration profile of camphor molecules on water.
With an increase in the surface-concentration, the surface tension 
decreases around the disk.
However, the force balance is kept as long as the surface
concentration profile is isotropic, and the disk is still stationary.

\item[Perturbation]
Small perturbations, such as the fluctuation of camphor concentration
or Brownian motion of the disk, break the symmetry of the concentration 
profile, resulting in an imbalance of surface tension acting on
the camphor disk.
Then, the disk experiences a small driving force toward
the lower concentration direction of camphor and
begins to migrate with low speed.

\item[Positive feedback]
The little shift of the disk position enhances the asymmetry of the 
camphor profile, and thus, the driving force of camphor motion
increases.
Owing to such a positive feedback process,
the initial small perturbation increases with time.

\item[Steady state]
Finally, the camphor disk achieves steady motion with a terminal velocity,
in which the driving force is balanced with the viscosity of water.
\end{description}

Nishimori et al.~\cite{NSN17} introduced the following
one-dimensional model:
Let $x_{\rm c}(t) \in \R$ be the position of the center of the camphor disk
at time $t \geq 0$ 
and define the velocity by $v_{\rm c}(t)=dx_{\rm c}(t)/dt$.
The surface-concentration of camphor is described by
a field $u(t, x)$, where $(t, x) \in [0, \infty) \times \R$ is
the spatio-temporal coordinate.
Then the following coupled system of
ordinary and
partial differential equations was considered; 
\begin{align}
m \frac{d v_{\rm c}(t)}{dt}
&= - \mu v_{\rm c}(t)
+ \frac{1}{2r} \Big[
\gamma(u(t, x_{\rm c}+r)) - \gamma(u(t,x_{\rm c}(t)-r)) \Big],
\label{eq:Eq1}
\\
\frac{\partial u(t,x)}{\partial t}
&=D \frac{\partial^2 u(t,x)}{\partial x^2}
- \kappa u(t,x) + S(x, x_{\rm c}(t); r),
\label{eq:Eq2}
\end{align}
where 
\begin{align}
\mu &=\mbox{the friction constant},
\nonumber\\
\gamma(u) &= \mbox{the surface tension
as a function of the camphor concentration field $u$},
\nonumber\\
r &=\mbox{the radius of the camphor disk},
\nonumber\\
D &=\mbox{effective diffusion constant of camphor on the water surface},
\nonumber\\
\kappa &=\mbox{the sum of sublimation constant
and dissolution}.
\nonumber\\
S(x, x_{\rm c}(t), r) &=
\mbox{the function of the supply rate of camphor from the disk at $x_{\rm c}(t)$}.
\label{eq:constants}
\end{align}
Nishimori et al.~\cite{NSN17} assumed the following
for $\gamma(u)$, 
\begin{equation}
\gamma(u) = \frac{\gamma_0-\gamma_1}{(\beta u)^2+1} + \gamma_1,
\label{eq:gamma}
\end{equation}
where $\beta$ is a positive constant,
$\gamma_0$ and $\gamma_1$ are the surface tensions of
pure water and saturated camphor aqueous solution respectively,
and
\begin{equation}
S(x, x_{\rm c}; r)
=\begin{cases}
S_0, \quad & |x-x_{\rm c}| \leq r,
\\
0, \quad & |x-x_{\rm c}| >r,
\end{cases}
\label{eq:S}
\end{equation}
where $S_0$ is the supply rate from the camphor grain
to the surface of water.
They also used another form for $S(x, x_{\rm c}; r)$ using the
Dirac delta function;
\begin{equation}
S(x, x_{\rm c}; r)=2 r S_0 \delta(x-x_{\rm c}).
\label{eq:S2}
\end{equation}

Nishimori et al.~\cite{NSN17}
have assumed that the relaxation rate of
$u(t, x)$ in \eqref{eq:Eq2} is much faster than the acceleration 
(deceleration) rate of $x_{\rm c}(t)$ in \eqref{eq:Eq1}. 
They have considered the case such that
the field of camphor surface-concentration $u(t, x)$
relaxes with negligibly short time to the
quasi-steady migration state following the quasi-steady
movement of the camphor disk with velocity $v_{\rm c}(t)$.

\SSC
{Stochastic Newtonian-Motion model}
\label{sec:newton_model}
\subsection{Newtonian equation coupled with random walks}
\label{sec:RW}

In the model system proposed by Nishimori et al.~\cite{NSN17},
the camphor disk is considered as a macroscopic
particle obeying the Newtonian equation \eqref{eq:Eq1},
while the field of camphor surface-concentration 
$u(t,x)$ follows the diffusion-type equation \eqref{eq:Eq2}.

In the following, we consider a microscopic model
for supply, diffusion, and sublimation
of camphor molecules on a surface of water
by using symmetric random walks which approximate
the Brownian motions of camphor molecules in water.
We consider the three-dimensional space 
$\R^3 \ni (x, y, z)$, where
the $z=0$ plane is regarded as the water surface.
We represent a camphor disk by a point mass
having mass $m$, which can move on the $z=0$ plane.
The random walks representing the camphor molecules
in water are moving only in the region $z \leq 0$.

\begin{description}
\item{(1)} \quad
We consider a discrete-time model 
with time $t \in \N_0:=\{1,2,\dots \}$.
The particle (camphor disk) 
releases $N$ random walkers 
(camphor molecule) 
every time $t$. 
In the following simulation, 
we set $N=50$.

This represents the last term, $S(x, x_{\rm c}(t); r)$, 
in the right-hand side of the diffusion-type equation
\eqref{eq:Eq2}.

\item{(2)} \quad
Each random walker independently performs
three-dimensional symmetric random walk 
in the region $z \leq 0$.
Set $m_1 \in \N:=\{1,2, \dots\}$ and let
the displacement of each hopping be
$1/\sqrt{m_1}$.
During the time unit, 
$m_2$ steps are performed 
by each random walker.
In the following simulation, 
we set $m_1=m_2=5$.

This represents the first term, 
$D \partial^2 u(t, x)/\partial x^2$, 
in the right-hand side of \eqref{eq:Eq2}.

\item{(3)} \quad
Set $q \in [0, 1]$.
If a random walker arrives at the $z=0$ plane,
and if it chooses the positive $z$-direction for a hopping,
the hopping is canceled and it remains at the same position in $z=0$
with probability $1-q$, and
it is annihilated with probability $q$.
In the following simulation,
we set $q=0.3$.

This represents the second term, $-\kappa u(t, x)$,
in the right-hand side of \eqref{eq:Eq2}
representing sublimation.
\end{description}

The position of the particle at time $t$
is denoted by $\r(t)$.
We write the coordinate at time $t$
of the $j$-th random walker ($j=1,2, \dots, N$)
starting from $\r(s)$ at time $s$
as $\W_j^{\r(s)}(t-s)$, 
$t =\{s, s+1, \dots\}$.
The time when it is annihilated at the 
$z=0$ plane is denoted by
$\tau_j^{\r(s)} \in \{s+1, s+2, \dots, \infty \}$.
We think that after this time, 
the random walk is to be the null state $\emptyset$;
$\W_j^{\r(s)}(t) \equiv \emptyset$
for $t \geq \tau_j^{\r(s)}$.
The empirical measure of random walks 
at time $t$ is denoted as
\begin{equation}
\Xi_t(\cdot) = 
\sum_{s=0}^{t-1}
\sum_{j=1}^{N}
\1(\tau_j^{\r(s)} > t)
\delta_{\W_j^{\r(s)}(m_2(t-s))/\sqrt{m_1}}(\cdot),
\label{eq:Xi}
\end{equation}
where 
$\1(\omega)$ denote the indicator function of 
a condition $\omega$:
$\1(\omega)=1$ if $\omega$ is satisfied,
$\1(\omega)=0$ otherwise. 
Here $\delta_{\w}(\cdot)$ is the point-mass (Dirac) measure
located at the position $\w$, and
for any subset $\Omega \subset \R^3$, the equality
\begin{equation}
\int_{\Omega} \delta_{\w}(d \r)=
\1(\w \in \Omega)
\label{eq:delta}
\end{equation}
is satisfied,
where 
$d \r=dx dy dz$. 

\begin{description}
\item{(4)} \quad
Assume that the position of the 
camphor disk is $\r=(x, y, z)$. 
We consider the rectangular parallelepiped regions
with sizes
\begin{equation}
L_x=2, \quad
L_y=2, \quad
L_z=1
\label{eq:size}
\end{equation}
in the $x$-, $y$-, and $z$-direction, respectively.
We set the following four regions around $\r$,
\begin{align}
\Lambda_{x+}(\r) &= \{(\xi_1, \xi_2, \xi_3): x \leq \xi_1 < x+L_x, 
y-L_y/2 \leq \xi_2 \leq y+L_y/2, 
-L_z \leq \xi_3 \leq 0\}, 
\nonumber\\
\Lambda_{x-}(\r) &= \{(\xi_1, \xi_2, \xi_3): x-L_x \leq \xi_1 < x, 
y-L_y/2 \leq \xi_2 \leq y+L_y/2, 
-L_z \leq \xi_3 \leq 0\}, 
\nonumber\\
\Lambda_{y+}(\r) &= \{(\xi_1, \xi_2, \xi_3): x-L_x/2 \leq \xi_1 \leq x+L_x/2, 
y \leq \xi_2 < y+L_y, 
-L_z \leq \xi_3 \leq 0\}, 
\nonumber\\
\Lambda_{y-}(\r) &= \{(\xi_1, \xi_2, \xi_3): x-L_x/2 \leq \xi_1 \leq x+L_x/2, 
y-L_y \leq \xi_2 < y, 
-L_z \leq \xi_3 \leq 0\},
\label{eq:Lambda}
\end{align}
and the union of them is written by
\begin{equation}
\Lambda(\r) = \Lambda_{x+}(\r) \cup \Lambda_{x-}(\r)
\cup \Lambda_{y+}(\r) \cup \Lambda_{y-}(\r).
\label{eq:LambdaB}
\end{equation}
For $\Xi_{t}$, the number of random walkers included
in each region is given by
\begin{align}
\Xi_{t}(\Lambda_{\sharp}(\r))
&= \int_{\Lambda_{\sharp}(\r)} \Xi_{t}(d \r')
\nonumber\\
&= \int_{\Lambda_{\sharp}(\r)} 
\sum_{s=0}^{t-1}
\sum_{j=1}^{N}
\1(\tau_j^{\r(s)} > t)
\delta_{\W_j^{\r(s)}(m_2(t-s))/\sqrt{m_1}}(d \r')
\nonumber\\
&= \sum_{s=0}^{t-1}
\sum_{j=1}^{N}
\1(\tau_j^{\r(s)} > t)
\1(
\W_j^{\r(s)}(m_2(t-s))/\sqrt{m_1}
\in \Lambda_{\sharp}(\r)),
\nonumber\\
& \qquad \qquad 
\sharp \in \{ x_+, x_-, y_+, y_- \}. 
\label{eq:number}
\end{align}
Due to the asymmetry of
the number of random walkers around the camphor disk,
the driving force acts the camphor disk toward
the direction of 
the smaller number of random walkers; 
\begin{align}
&\F(\r(t), \Xi_t)=(F_x(\r(t), \Xi_t), 
F_y(\r(t), \Xi_t)), 
\nonumber\\
& \qquad 
F_x(\r(t), \Xi_t)
=- \beta \Big[ \Xi_t(\Lambda_{x+}(\r(t)))
- \Xi_t(\Lambda_{x-}(\r(t))) 
\Big], 
\nonumber\\
& \qquad 
F_y(\r(t), \Xi_t)
=- \beta \Big[ \Xi_t(\Lambda_{y+}
(\r(t)))- \Xi_t(\Lambda_{y-}(\r(t))) 
\Big], \quad t \in \N_0, 
\label{eq:F}
\end{align}
where $\beta$ is a positive constant. 

This represents the second term, 
$[\gamma(u(t, x_{\rm c}+r))-r(u(t, x_{\rm c}-r)]/2r$,
in the right-hand side of the Newtonian equation \eqref{eq:Eq1}.

\item{(5)} \quad
Set the initial state of the camphor disk;
\begin{equation}
\r(0): \mbox{initial position}, 
\quad 
\v(0) : \mbox{initial velocity}.
\label{eq:initial}
\end{equation}
Notice that both are vectors on the $z=0$ plane
representing the water surface.

\item{(6)} \quad
For each discrete time $t \in \N_0$, 
evolve the position and velocity of camphor disk
by the following difference equations,
\begin{align}
m\Big[ 
\v(t+1)-\v(t) \Big] &=-\mu \v(t)+ \F(\r(t), \Xi_t),
\nonumber\\
\r(t+1)& = \r(t) + \v(t+1).
\label{eq:eq_motion}
\end{align}
\end{description}
\noindent{\bf Remark 1} \,
In the original model by Nishimori et al.~\cite{NSN17}, 
the surface tension $\gamma$ changes depending on
the surface-concentration of camphor $u$ 
by the formula \eqref{eq:gamma},
and the spatial asymmetry of surface tension 
causes the driving force of camphor disk
as given by the second term of the right-hand side
of \eqref{eq:Eq1}.
Here we consider a linearization of the surface 
tension $\gamma$
of $u$ and then the driving force is assumed to be simply
proportional to the spatial asymmetry of 
the camphor surface-concentration $u(t, x)$ as \eqref{eq:F}. 

\subsection{Simulation Results}
\label{sec:simulation}
\subsubsection{Single camphor disk}
\label{sec:one_float}

\begin{figure}[ht]
\begin{center}
\includegraphics[width=0.5\textwidth]{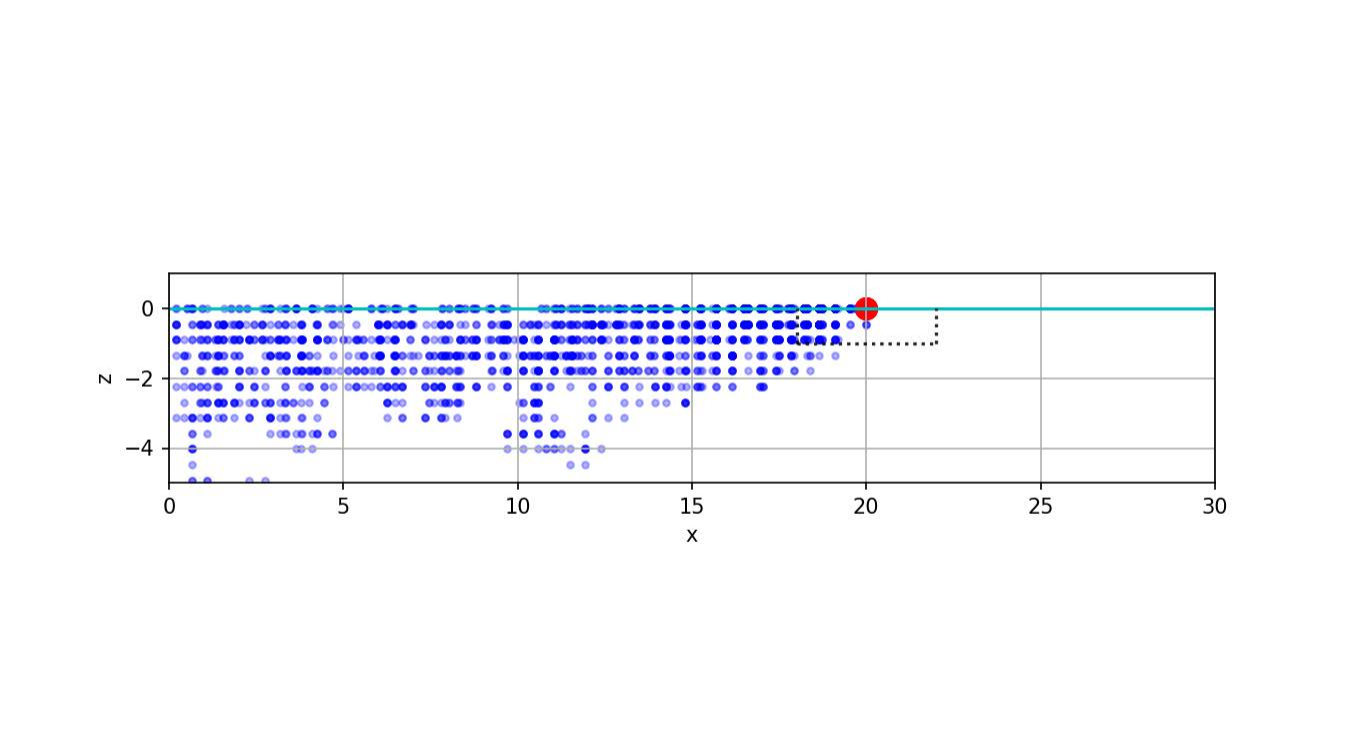}
\end{center}
\vskip -1.5cm
\caption{Single camphor disk (red disk) and 
random walkers representing camphor molecules (blue dots)
in the $(x, z)$-plane.
The dotted rectangular around the red disk
indicates $\Lambda(\r)$.}
\label{fig:1_xz} 
\begin{center}
\includegraphics[width=0.5\textwidth]{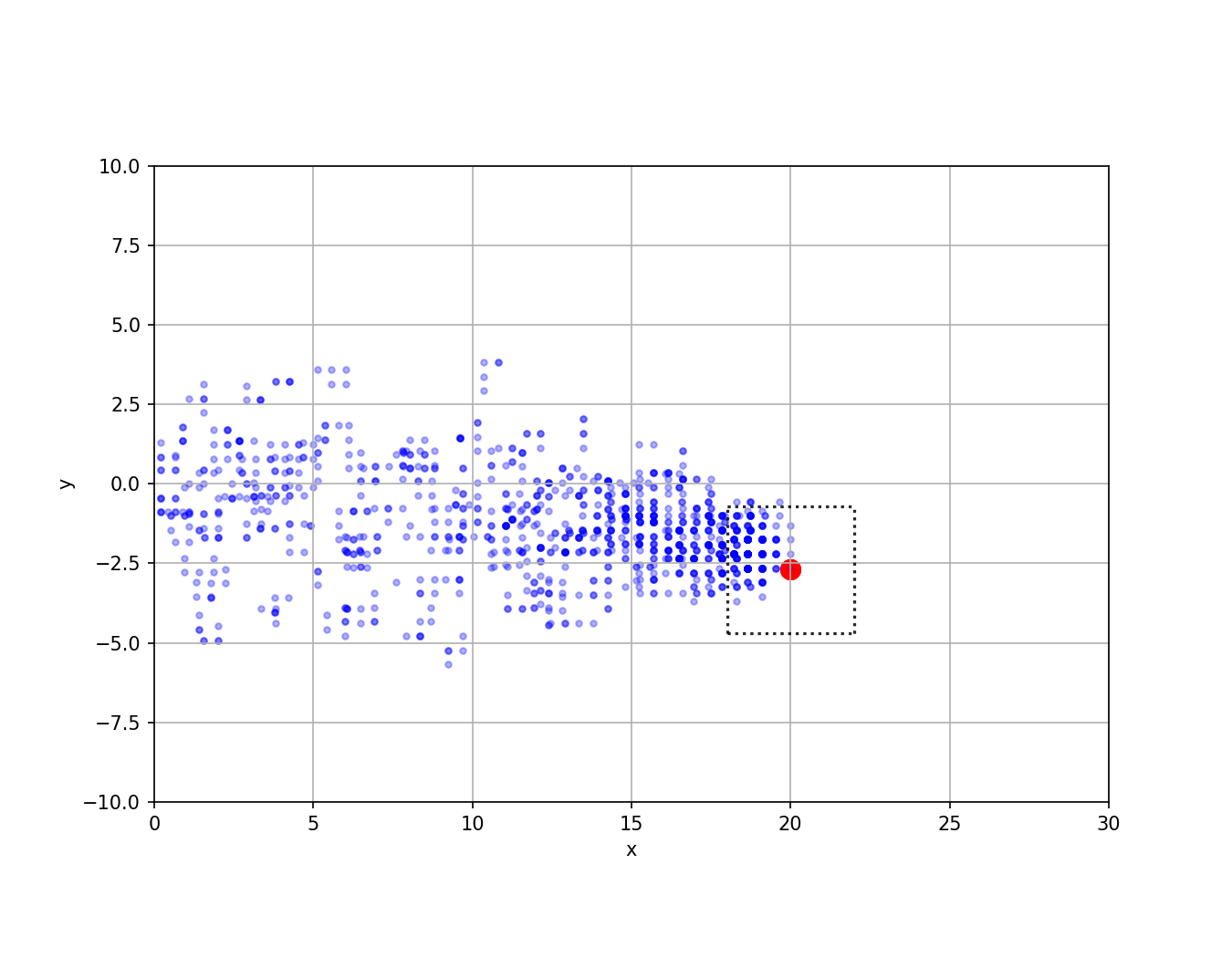}
\end{center}
\vskip -1.2cm
\caption{Single camphor disk (red disk) and 
random walkers representing camphor molecules (blue dots)
in the $(x, y)$-plane.
The dotted square around the red disk
indicates $\Lambda(\r)$.}
\label{fig:1_xy}
\label{fig:one}
\end{figure}

First we have simulated the motion of single camphor disk,
where we put the initial velocity in the positive $x$-direction, 
$\v(0)=(2, 0, 0)$ at $\r(0)=(0, 0, 0)$.
Figures~\ref{fig:1_xz} and \ref{fig:1_xy} 
show the position of camphor disk (red disk) 
and the distribution of random walkers representing
the camphor molecules (blue dots) a while later. 
Here we set $m_1=m_2=3$ and $\beta/\mu=0.05$.
Other parameters are the same as mentioned above.
As shown by figures, the density of random walkers
has higher value in the left (backward) than 
in the right (forward) of the camphor disk.
Hence the driving force of camphor disk
is in the right direction and
the disk will continue its motion to the right.

By the density fluctuation of random walkers,
the velocity of camphor disk becomes to have
$y$-component as shown by Fig.~\ref{fig:1_xy}.

\subsubsection{Two camphor disks}
\label{sec:two_float}

\begin{figure}[ht]
\begin{center}
\includegraphics[width=0.5\textwidth]{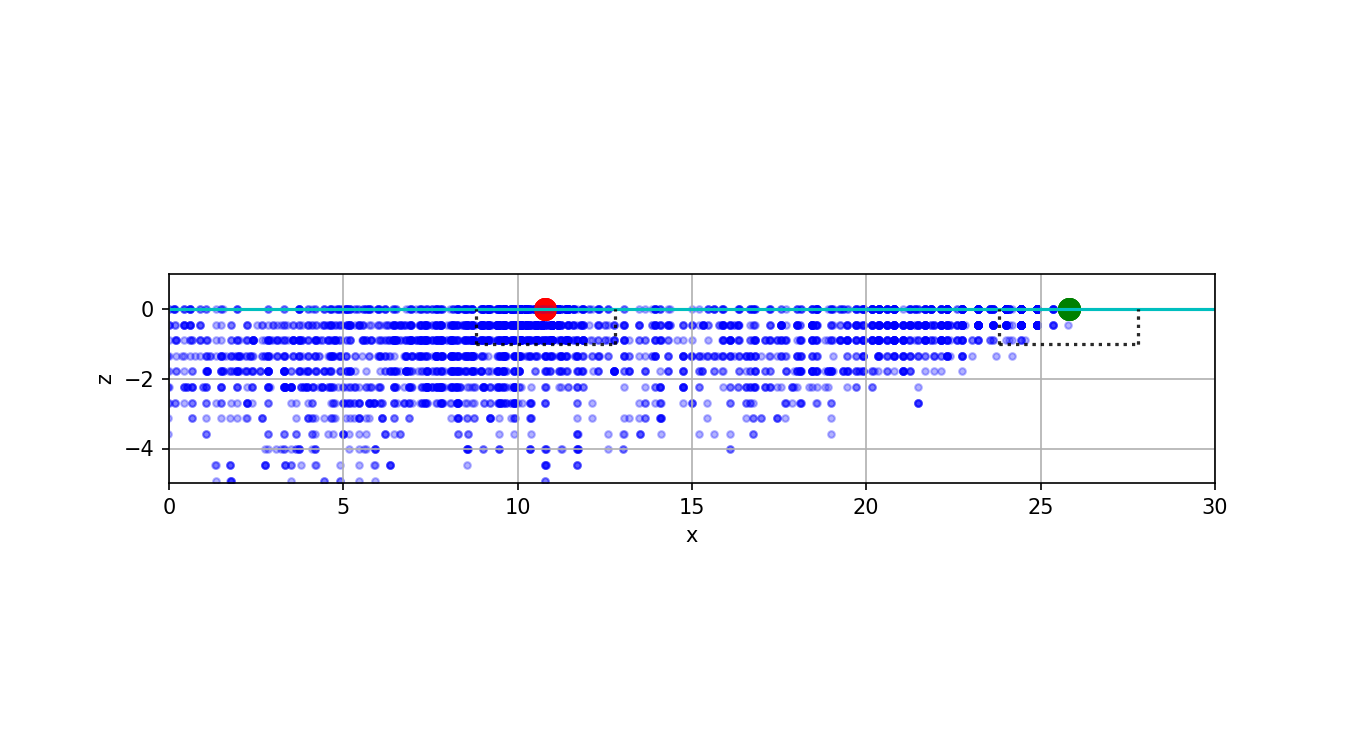}
\end{center}
\vskip -1.5cm
\caption{Two camphor disks (red and green disks) and 
random walkers representing camphor molecules (blue dots)
in the $(x, z)$-plane.
}
\label{fig:2_xz}
\begin{center}
\includegraphics[width=0.5\textwidth]{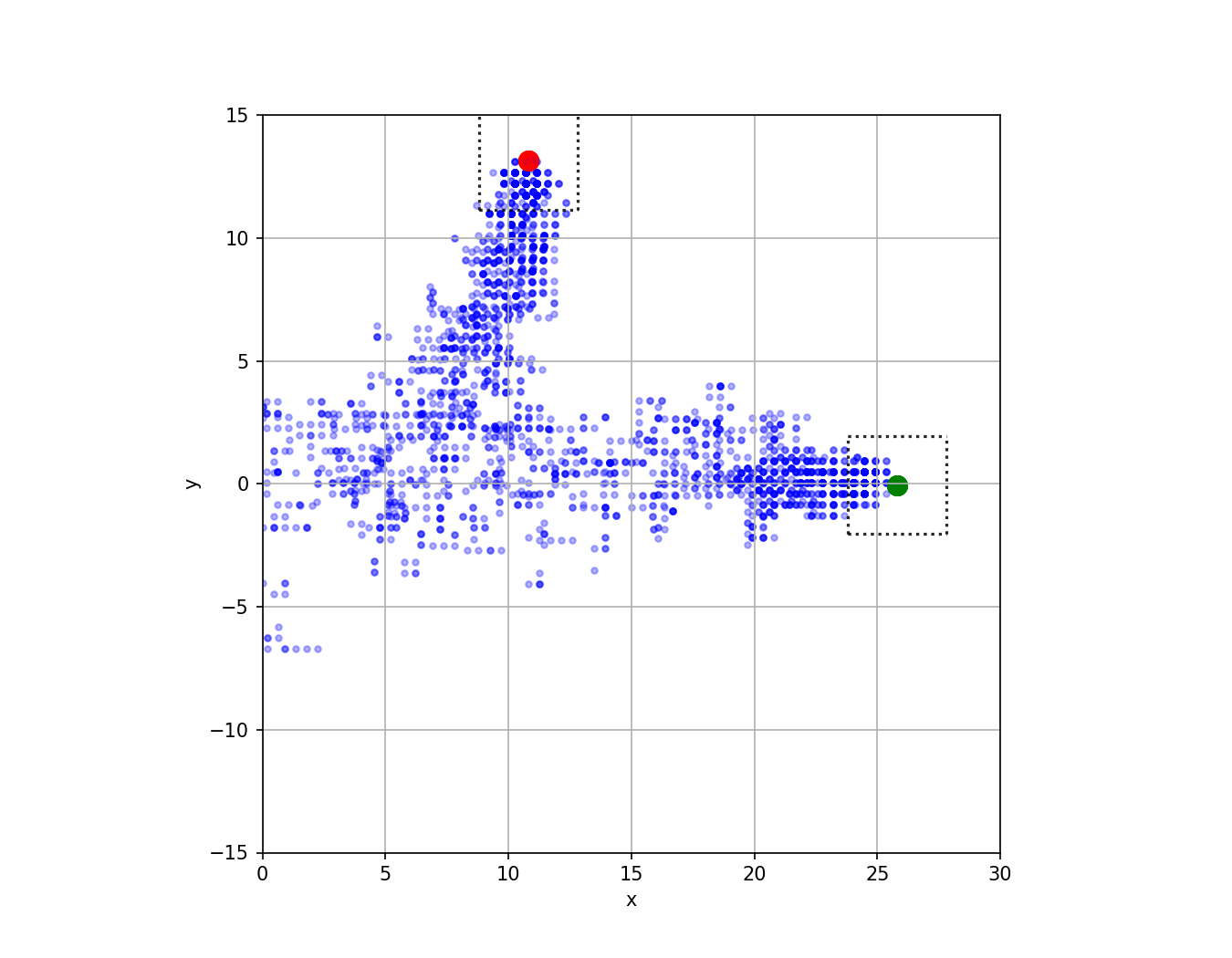}
\end{center}
\vskip -1cm
\caption{Two camphor disks (red and green disks) and 
random walkers representing camphor molecules (blue dots)
in the $(x, y)$-plane.}
\label{fig:2_xy}
\end{figure}

Next we put two camphor disks in the $x$-axis
with a small distance;
$\r_1(0)=(0, 0, 0)$ and $\r_2(0)=(5, 0, 0)$.
We set the initial velocities as 
$\v_1(0)=\v_2(0)=(2, 0, 0)$. 
Figures~\ref{fig:2_xz} and \ref{fig:2_xy} 
show the positions of two camphor disks
(red and green disks)
and distribution of random walkers (blue dots) a while later.
We can observe a repulsive behavior of two disks,
and then they move to the different directions as
shown by Fig.~\ref{fig:2_xy}. 

\subsubsection{Three and four camphor disks}
\label{sec:3_4_disk}

We set three (resp. four) camphor disks to make
an equilateral triangle (resp. a square).
We put appropriate initial velocities to make them
collide at the center of the triangle or square.
We have observed that the camphor disks
show repulsive interaction 
and they are scattered from each other.

\SSC
{Stochastic Viscous-Motion Models and Dynamical Model
in One-Dimension}
\label{sec:viscous}

Here we consider the one-dimensional case
in order to simplify the analysis
and clarify the essential aspects of the present modeling.

\subsection{Stochastic models}
\label{sec:stochastic_viscous}

When the friction constant $\mu$ is large,
acceleration is suppressed and
the left-hand side of \eqref{eq:eq_motion}
will be negligible.
In such a case dominated by viscosity, 
we consider the system
\begin{align}
v(t+1)
&=\frac{1}{\mu} F(x(t), \Xi_t),
\label{eq:viscous1a}
\\
x(t+1) &= x(t) + v(t+1),
\quad
t \in \N_0,
\label{eq:viscous1b}
\end{align}
which we call
the \textit{stochastic viscous-motion model}.
Equation \eqref{eq:viscous1a} is written as
\begin{align}
&v(t+1)
=-\frac{\beta}{\mu} 
\Big[ \Xi_t(\Lambda_{x+}(x(t))
- \Xi_t (\Lambda_{x-}(x(t))
\Big]
\nonumber\\
\iff \,
&v(t+1)
\nonumber\\
& \,
=-\frac{\beta}{\mu} \sum_{s=0}^{t-1}
\left[
\sum_{j=1}^{N} 
\1(\tau_j^{x(s)} > t)
\1\left( 
\frac{1}{\sqrt{m_1}} W_j^{x(s)}(m_2(t-s)) 
\in [x(t), x(t)+L_x) \right)
\right.
\nonumber\\
& \qquad \qquad
\left.
-
\sum_{j=1}^{N} 
\1(\tau_j^{x(s)} > t)
\1\left( 
\frac{1}{\sqrt{m_1}} W_j^{x(s)}(m_2(t-s))
\in [x(t)-L_x, x(t)) \right) \right], 
\label{eq:viscous2}
\end{align}
$t \in \N_0$.
We will consider the 
simple case
\begin{equation}
\1(\tau_j^{x(s)} > t)
=\begin{cases}
1, \quad & \mbox{if $s=t-1$}, \cr
0, \quad & \mbox{otherwise}.
\end{cases}
\label{eq:simple}
\end{equation}
That is, the random walkers survive 
(the camphor molecules stay in water
avoiding from sublimation to the air)
only for the time unit)
In this case, we have
\begin{align}
v(t+1)
&
=-\frac{\beta}{\mu} 
\left[
\sum_{j=1}^{N} 
\1\left( 
\frac{1}{\sqrt{m_1}}
W_j^{x(t-1)}(m_2)
\in [x(t), x(t)+L_x) \right)
\right.
\nonumber\\
& \qquad \quad
\left.
-
\sum_{j=1}^{N} 
\1\left( 
\frac{1}{\sqrt{m_1}}
W_j^{x(t-1)}(m_2)
\in [x(t)-L_x, x(t)) \right) \right], 
\nonumber\\
x(t+1) &=x(t)+v(t+1),
\quad t \in \N_0.
\label{eq:simple_viscous}
\end{align}

\subsubsection{Brownian motion limit}
\label{eq:BM}
In the following limit, the scaled random walk
converges to the one-dimensional standard Brownian 
motion in probability; 
for a given initial position $x \in \R$, 
\begin{equation}
\left( \frac{1}{\sqrt{m_1}} W^{x}(m_2 t) \right)_{t \geq 0}
\longrightarrow 
\Big( \sqrt{2D} B^x(t) \Big)_{t \geq 0} \quad
\mbox{as} \quad
m_1, m_2 \to \infty \quad
\mbox{with} \quad
\frac{m_2}{m_1} \to 2 D,
\label{eq:BM1}
\end{equation}
where $(B^x(t))_{t \geq 0}$ denotes the one-dimensional 
standard Brownian motion started from $x$.
In such a \textit{diffusion scaling limit}, 
Eq.~\eqref{eq:viscous2} is reduced to
\begin{align}
v(t+1)
&
=-\frac{\beta}{\mu}
\sum_{s=0}^{t-1}
\left[
\sum_{j=1}^{N} 
\1(\tau_j^{x(s)} > t)
\1\left( 
\sqrt{2D}
B_j^{x(s)}(t-s)
\in [x(t), x(t)+ L_x) \right)
\right.
\nonumber\\
& \qquad \qquad
\left.
-
\sum_{j=1}^{N} 
\1(\tau_j^{x(s)} > t)
\1\left( 
\sqrt{2D}
B_j^{x(s)}(t-s)
\in [x(t)-L_x, x(t)) \right)
\right], \quad t \in \N_0.
\label{eq:BM3}
\end{align}
Here $\tau_j^{x(s)}$ denote
the time that the limit Brownian motion 
$(B^{x(s)}(t))_{t \geq 0}$ is annihilated. 
The position of camphor disk is moving as
\begin{equation}
x(t+1)=x(t)+v(t+1),
\quad t \in \N_0.
\label{eq:BM4}
\end{equation}

\subsubsection{Mean-value motion}
\label{eq:mv}

Now we take the mean values with respect to
the Brownian motions,
\begin{align}
\bra v(t+1) \ket
&
=-\frac{\beta}{\mu}
\sum_{s=0}^{t-1}
\left[
\sum_{j=1}^{N} 
\left\bra
\1(\tau_j^{x(s)} > t)
\1\left( 
\sqrt{2D}
B_j^{x(s)}(t-s)
\in [x(t), x(t)+L_x) \right)
\right\ket
\right.
\nonumber\\
& \qquad \qquad
\left.
-
\sum_{j=1}^{N} 
\left\bra
\1(\tau_j^{x(s)} > t)
\1\left( 
\sqrt{2D}
B_j^{x(s)}(t-s) \in [x(t)-L_x, x(t)) \right)
\right\ket
\right],
\label{eq:mv1}
\end{align}
The transition probability density of the
one-dimensional standard Brownian motion
is given by
\begin{equation}
p(t, y|x)=\frac{1}{\sqrt{2 \pi t}}
e^{-(y-x)^2/2t},
\quad t >0, \quad x, y \in \R.
\label{eq:tpd}
\end{equation}
Then, provided $x(s) \in \R$, $s=0, 1, \dots, t-1$, 
the right-hand side of 
\eqref{eq:mv1} will be written as
\begin{align}
&
- N \frac{\beta}{\mu} \sum_{s=0}^{t-1}
G(t-s)
\left[ \int_{x(t)}^{x(t)+L_x} p(2D(t-s), x|x(s)) dx
-\int_{x(t)-L_x}^{x(t)} p(2D(t-s), x|x(s)) dx
\right], 
\label{eq:cal1}
\end{align}
where we have assumed independence of
the annihilation process of Brownian motion
and its position, and defined the function
\begin{equation}
G(t-s) := \bra \1( \tau^{x(s)} > t) \ket,
\quad t > s,
\label{eq:G}
\end{equation}
which corresponds to the survival probability of
the camphor molecule in water
avoiding from sublimation for time period $t-s$.
We cal \eqref{eq:G} the
\textit{sublimation function}.
If $L_x \ll 1$, the above will be written as
\begin{align}
& - N \frac{\beta L_x}{\mu} N
\sum_{s=1}^{t-1} G(t-s)
\Big[ 
p(2D(t-s), x(t)+L_x/2|x(s)) 
-
p(2D(t-s), x(t)-L_x/2|x(s)) \Big]
\nonumber\\
& \hskip 5cm
+{\rm O}(L_x^2)
\nonumber\\
& 
=- N \frac{\beta L_x^2}{\mu} 
\sum_{s=1}^{t-1} G(t-s)
\frac{
p(2D(t-s), x(t)+L_x/2|x(s)) 
-
p(2D(t-s), x(t)-L_x/2|x(s))}{L_x}
\nonumber\\
& \hskip 5cm
+{\rm O}(L_x^2)
\label{eq:cal2}
\end{align}
We take the following scaling limit,
\begin{equation}
N \to \infty, \quad 
L_x \to 0 \quad
\mbox{with} \quad 
N \frac{\beta L_x^2}{\mu} \to \alpha \ell, 
\label{eq:scaling_limit}
\end{equation}
where $0< \alpha \ell < \infty$.
Then \eqref{eq:cal2} will converge to 
\begin{equation}
-\alpha \ell \sum_{s=1}^{t-1} G(t-s)
\frac{\partial}{\partial x} 
p(2D(t-s), x| x(s)) \Big|_{x=x(t)}.
\label{eq:scaling_limit2}
\end{equation}
The above calculation implies the
\textit{Markov chain}
\begin{equation}
(x(t), v(t)) \to
(x(t+1), v(t+1)),
\quad t \in \N_0.
\label{eq:Markov1}
\end{equation}
with 
\begin{align}
v(t+1) &=- \alpha \ell 
\sum_{s=1}^{t-1} G(t-s) 
\frac{\partial}{\partial x} 
p(2D(t-s), x| x(s))
\Big|_{x=x(t)},
\nonumber\\
x(t+1) &=x(t)+v(t+1).
\label{eq:Markov2}
\end{align}
Two examples of sublimation function $G$
are given by
\begin{equation}
G(u)=e^{-\kappa u}, \quad u >0.
\label{eq:G1}
\end{equation}
and
\begin{equation}
G(u) = \begin{cases}
1, \quad &\mbox{if $u=1$},
\\
0, \quad & \mbox{if $u >1$}.
\end{cases}
\label{eq:G2}
\end{equation}
The former corresponds to the second term,
$-\kappa u(t, x)$, in the right-hand side of
\eqref{eq:Eq1} of the model by Nishimori et al.~\cite{NSN17}.

\subsection{Dynamical system}
\label{sec:dynamical}

From now on, we will consider the simple case
with \eqref{eq:G2} for the sublimation function.
Then \eqref{eq:Markov2} gives the
following deterministic system with
discrete-time, which we simply call
the \textit{dynamical system}, 
\begin{align}
v(t+1) &=- \alpha \ell 
\frac{\partial}{\partial x} 
p(2D, x| x(k))
\Big|_{x=x(t)},
\label{eq:dynamical1}
\\
x(t+1) &=x(t)+v(t+1),
\quad t \in \N_0.
\label{eq:dynamical2}
\end{align}
The right-hand side of \eqref{eq:dynamical1} is calculated as
\begin{align*}
&\left. 
-\alpha \ell \frac{\partial}{\partial x}
\frac{1}{\sqrt{4 \pi D}} e^{-(x-x(n-1))^2/4D} 
\right|_{x=x(n)}
\nonumber\\
& \quad
=\left. -\frac{\alpha \ell}{\sqrt{4 \pi D}}
\left( - \frac{2(x-x(n-1))}{4D} \right)
e^{-(x-x(n-1))^2/4D} 
\right|_{x=x(n)}
\nonumber\\
& \quad
=\frac{\alpha \ell}{4 \sqrt{\pi} D^{3/2}}
(x(n)-x(n-1)) e^{-(x(n)-x(n-1))^2/4D}.
\end{align*}
By \eqref{eq:dynamical2},
\[
x(n)-x(n-1)=v(n).
\]
Then the time-evolution equation of velocity
is given by the following iterative equation,
\begin{equation}
v(n+1)= \frac{\alpha \ell}{4 \sqrt{\pi} D^{3/2}}
v(n) e^{-v(n)^2/4D},
\quad n \in \N_0.
\label{eq:dynamics3}
\end{equation}

\subsubsection{Analysis of dynamical system}
\label{sec:analysis}

Set
\begin{equation}
V(n):= \frac{v(n)}{2 \sqrt{D}}, \quad n \in \N_0,
\label{eq:V1}
\end{equation}
and
\begin{equation}
C:=\frac{4 \sqrt{\pi} D^{3/2}}{\alpha \ell} >0.
\label{eq:C}
\end{equation}
Then \eqref{eq:dynamics3} is written as
\begin{equation}
\frac{v(n+1)}{2 \sqrt{D}}
= \frac{\alpha \ell}{4 \sqrt{\pi} D^{3/2}} 
\frac{v(n)}{2 \sqrt{D}} e^{-(v(n)/2 \sqrt{D})^2}
\, \iff \,
C V(n+1)=F(V(n)), \quad n \in \N_0
\label{eq:Veq1}
\end{equation}
with
\begin{equation}
F(x) = x e^{-x^2}.
\label{eq:F1}
\end{equation}

\vskip 0.3cm
\noindent{\bf Remark 2} \, 
If we take the square of \eqref{eq:Veq1},
we obtain
\[
C^2 V(n+1) = V(n)^2 e^{-2V(n)^2}
\, \iff \,
-2 C^2 V(n+1) = -2 V(n)^2 e^{-2 V(n)^2}.
\]
Let
\begin{equation}
Y(n) :=-2 V(n)^2 \leq 0.
\label{eq:Y1}
\end{equation}
Then we have
\begin{equation}
C^2 Y(n+1)= Y(n) e^{Y(n)}, \quad n \in \N_0.
\label{eq:Y2}
\end{equation}
Here we consider the {\it Lambert $W$ function} 
(see, for instance, \cite{Veb12}).
This function is defined as the inverse function
of the mapping
\[
x \mapsto x e^{x}.
\]
This mapping is not injective, and the Lambert $W$
function has two branches with 
a branching point at $(-e^{-1}, -1)$ 
in the plane $(x, W) \in \R^2$.
The upper branch is denoted by $W_0(x)$ defined
for $x \in [-e^{-1}, \infty)$.
By this definition, we can show that
\[
W_0(x) e^{W_0(x)} = x,
\quad
W_0(0) = 0, \quad W_0(e)=1,
\]
and
$W_0(x) \simeq x$ as $x \to 0$.
Another branch is denoted by $W_{-1}(x)$.
See \cite{CGHJK96} for more details.
Using the Lambert $W$ function, \eqref{eq:Y2}
is written as
\begin{equation}
W(C^2 Y(n+1))=Y(n), \quad n \in \N_0.
\label{eq:equation5}
\end{equation}

\vskip 0.3cm
\noindent
\underline{\bf Fixed points}
\vskip 0.3cm
If we write the stationary solution of \eqref{eq:Veq1} as
$V_*=v_*/(2 \sqrt{D})$, then we have
\begin{equation}
C V_* = V_* e^{-V_*^2}.
\label{eq:stat1}
\end{equation}
It has a trivial solution
\begin{equation}
V_{*}=0 \quad \iff \quad v_{*}=0.
\label{eq:stat2}
\end{equation}
The non-zero solutions should satisfy
\begin{equation}
C=e^{-V_*^2}.
\label{eq:stat3}
\end{equation}
If and only if $C <1$, non-zero solution exists.
Assume $C < 1$. Then \eqref{eq:stat3} have solutions
\begin{align}
\log C=-V_*^2 \quad 
&\iff \quad
V_* = \pm \sqrt{-\log C}
\nonumber\\
&\iff \quad v^*=\pm 2 \sqrt{D} V^*
=\pm 2 \sqrt{-D \log C}.
\label{eq:stat4}
\end{align}

\begin{figure}[ht]
\begin{center}
\includegraphics[width=0.7\textwidth]{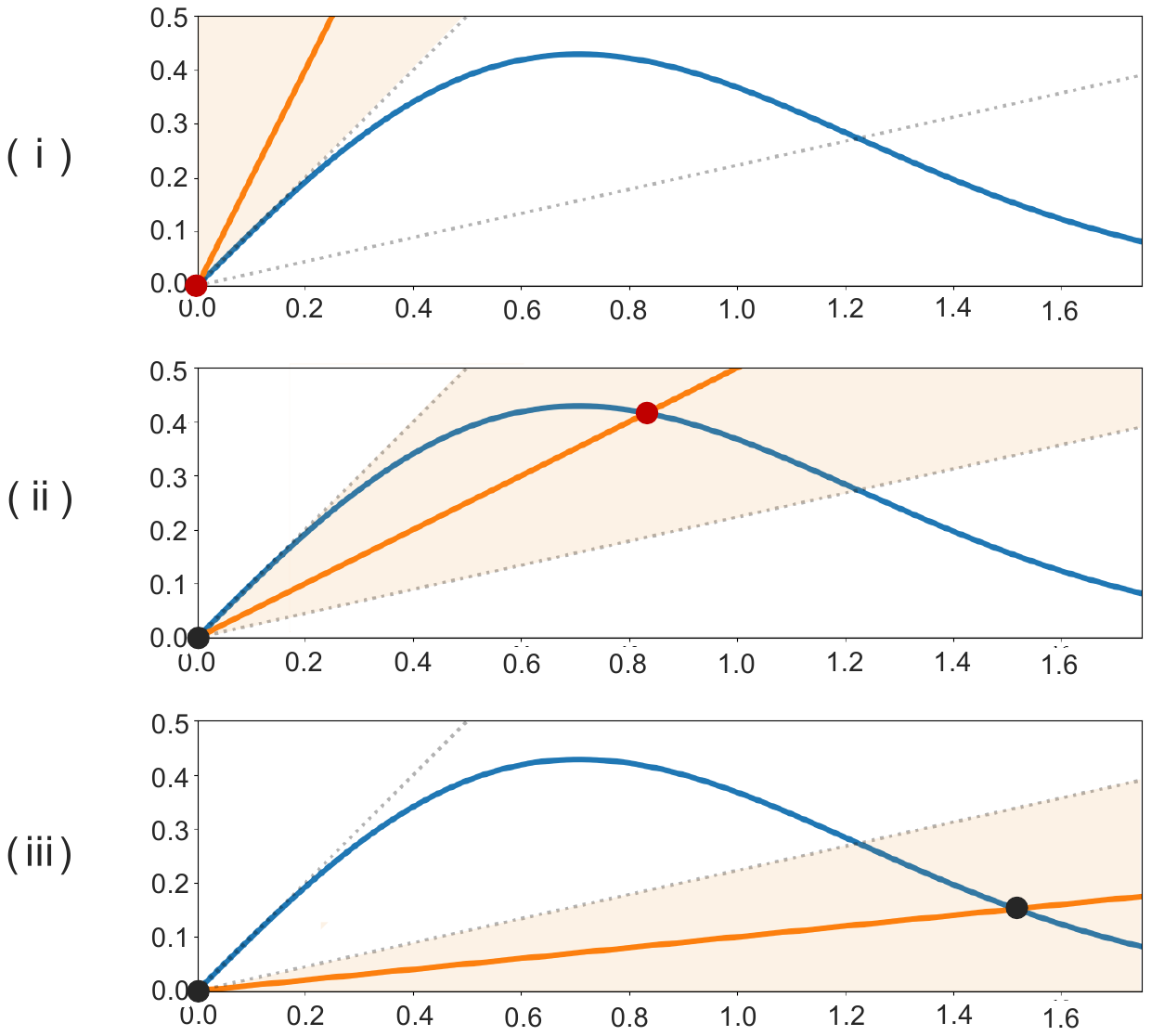}
\end{center}
\caption{The functions $y=F(x)$ given by \eqref{eq:F1}
and $y=C x$ are drawn 
by a blue curve and a red line, respectively
for three cases.
(i) $C \geq 1$, (ii) $e^{-1} < C < 1$, and
(iii) $0 < C < e^{-1}$. 
The stable (resp. unstable) fixed points
are indicated by red (resp. black) dots.}
\label{fig:stability}
\end{figure}

\vskip 0.3cm
\noindent
\underline{\bf Linear stability}
\vskip 0.3cm

We can prove the following by
the linear stability analysis.
See Fig.~\ref{fig:stability}.

\begin{prop}
\label{thm:stat}
\begin{description}
\item{\rm (i)} \quad
Assume 
\[
C \geq 1 \quad \iff \quad
\frac{4 \sqrt{\pi} D^{3/2}}{\alpha \ell} \geq 1.
\]
Then the stationary solution is uniquely given by
\[
V_* =0 \quad \iff \quad v_*=0.
\]
The zero solution $v_*=0$ 
(the rest state) is stable.
\item{\rm (ii)} \quad
Assume
\[
e^{-1}=0.3678... < C < 1
\quad \iff \quad
e^{-1} < 
\frac{4 \sqrt{\pi} D^{3/2}}{\alpha \ell} < 1.
\]
Then there are three stationary solutions
\[
V_* =0 \quad \mbox{and}
\quad \pm V_* =\pm \sqrt{-\log C}
\quad \iff \quad
v_* =0 \quad \mbox{and} 
\quad v_* = \pm 2 \sqrt{-D\log C}.
\]
The zero solution $v_*=0$ is unstable and
the non-zero solutions
$v_* = \pm 2 \sqrt{-D \log C}$
(the steady motions) are stable.
\item{\rm (iii)} \quad
Assume
\[
0 < C < e^{-1} \quad \iff \quad
0 < \frac{4 \sqrt{\pi} D^{3/2}}{\alpha \ell} < e^{-1}. 
\]
Then there are three stationary solutions
\[
V_* =0 \quad \mbox{and}
\quad \pm V_* = \pm \sqrt{-\log C}
\quad \iff \quad
v_* =0 \quad \mbox{and} 
\quad v_* = \pm 2 \sqrt{-D\log C}.
\]
Both of the zero solution $v_*=0$ 
and
the non-zero solutions
$v_* = \pm 2 \sqrt{-D \log C}$
are unstable
and oscillatory motion will be observed.
\end{description}
\end{prop}
\noindent{\it Proof} \quad
Let
\begin{equation}
V(n)=V_*+\varepsilon \quad
\mbox{with} \quad |\varepsilon| \ll 1.
\label{eq:Vt}
\end{equation}
Then by the equation of motion \eqref{eq:Veq1},
\begin{align*}
V(n+1) &=\frac{1}{C} V(n) e^{-V(n)^2}
\nonumber\\
&= \frac{1}{C} (V_*+\varepsilon) e^{-(V_*+\varepsilon)^2}
\nonumber\\
&=\frac{1}{C} V_* e^{-V_*^2}
+\frac{1}{C} e^{-V_*^2} 
(1-2 V_*^2 ) \varepsilon
+ {\rm O}(\varepsilon^2).
\end{align*}
By \eqref{eq:stat1}, we have 
\begin{equation}
V(n+1) =V_*
+\frac{1}{C} e^{-V_*^2} 
(1-2 V_*^2 ) \varepsilon
+ {\rm O}(\varepsilon^2).
\label{eq:error1}
\end{equation}
We define
\begin{equation}
\lambda := \lim_{\varepsilon \downarrow 0}
\frac{V(n+1)-V_*}{V(n)-V_*}.
\label{eq:lambda1}
\end{equation}
Then \eqref{eq:error1} gives
\begin{align}
\lambda &=\lim_{\varepsilon \downarrow 0}
\frac{1}{\varepsilon}
\left[
\frac{1}{C} e^{-V_*^2} 
(1-2 V_*^2 ) \varepsilon
+ {\rm O}(\varepsilon^2) \right]
=\frac{1}{C} e^{-V_*^2} 
(1-2 V_*^2 ).
\label{eq:lambda2}
\end{align}
By definition \eqref{eq:lambda1}, we can say that
for infinitesimal perturbation
\begin{align}
|\lambda| < 1 \quad
&\iff \quad \mbox{$V_*$ is stable},
\nonumber\\
|\lambda| > 1 \quad
&\iff \quad \mbox{$V_*$ is unstable}.
\label{eq:stability}
\end{align}
For $V_*=0$, \eqref{eq:lambda2} gives
\begin{equation}
\lambda=\frac{1}{C}.
\label{eq:lambda3}
\end{equation}
Then 
\begin{align}
C > 1 \quad
&\Longrightarrow \quad
\mbox{$V_*=0$ is stable},
\nonumber\\
0< C < 1 \quad
&\Longrightarrow \quad
\mbox{$V_*=0$ is unstable}.
\label{eq:stability0A}
\end{align}
For $V*=\pm \sqrt{- \log C}$,
\eqref{eq:lambda2} gives
\begin{equation}
\lambda=\frac{1}{C} \times C (1+2 \log C)
=1+ 2 \log C.
\label{eq:lambda4}
\end{equation}
Then
\begin{align}
e^{-1} < C < 1 \, \, 
&\iff \, \, 
-1 < \lambda < 1 \, \, 
\Longrightarrow
\quad \mbox{$V_*=\pm \sqrt{-\log C}$ is stable},
\nonumber\\
0 < C < e^{-1} \, \, 
&\iff \, \, 
\lambda < -1 \, \, 
\Longrightarrow \quad \mbox{$V_*=\pm \sqrt{-\log C}$ is unstable}.
\label{eq:stability1A}
\end{align}
Then the proof is complete. \qed


\subsection{Comparison between stochastic
models and dynamical systems}
\label{sec:comparison}
\begin{figure}[ht]
    \begin{tabular}{cc}
      \begin{minipage}[ht]{0.4\hsize}
        \centering
        \includegraphics[keepaspectratio, scale=0.25]{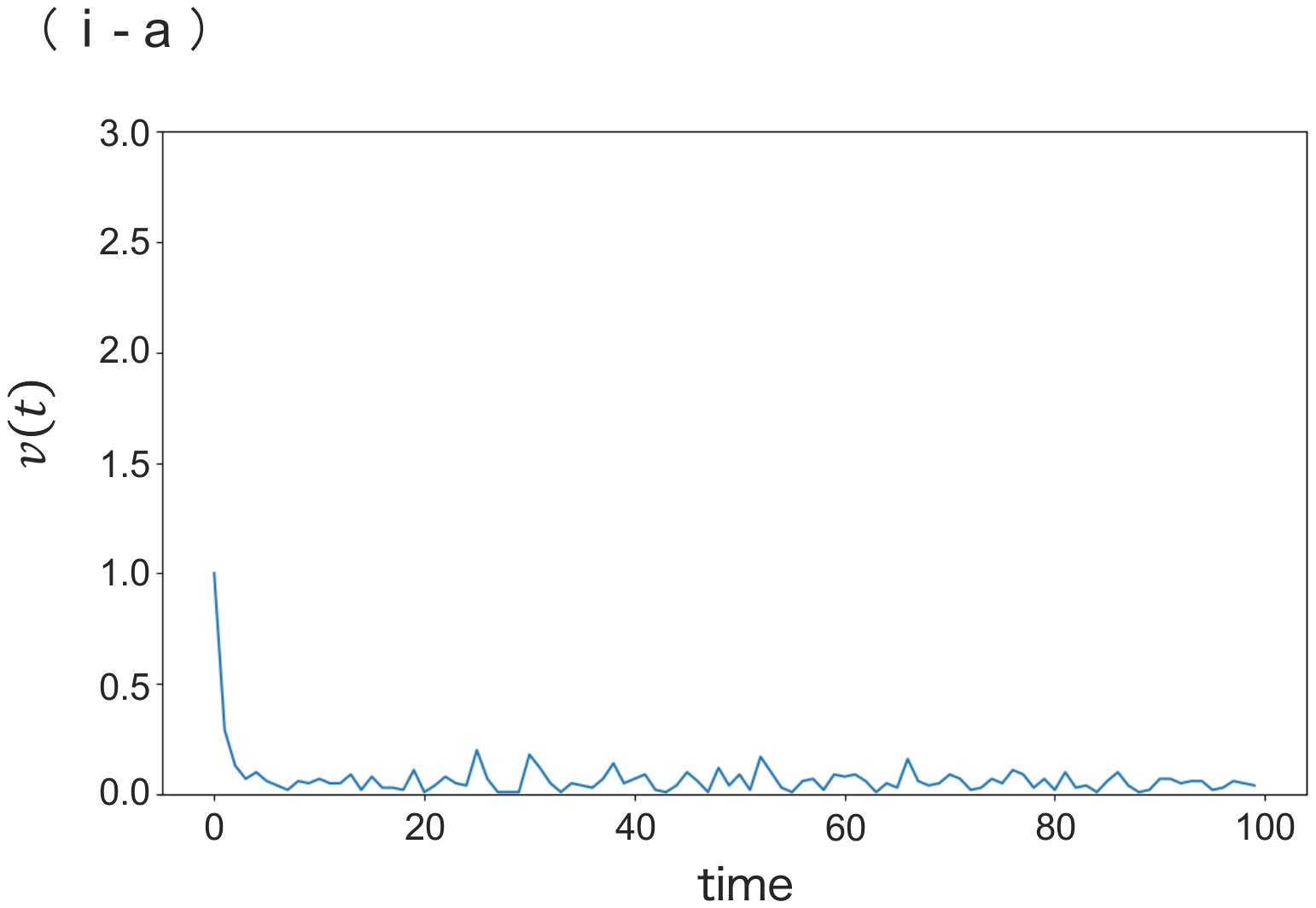}
      \end{minipage} &
      \begin{minipage}[ht]{0.4\hsize}
        \centering
         \includegraphics[keepaspectratio, scale=0.25]{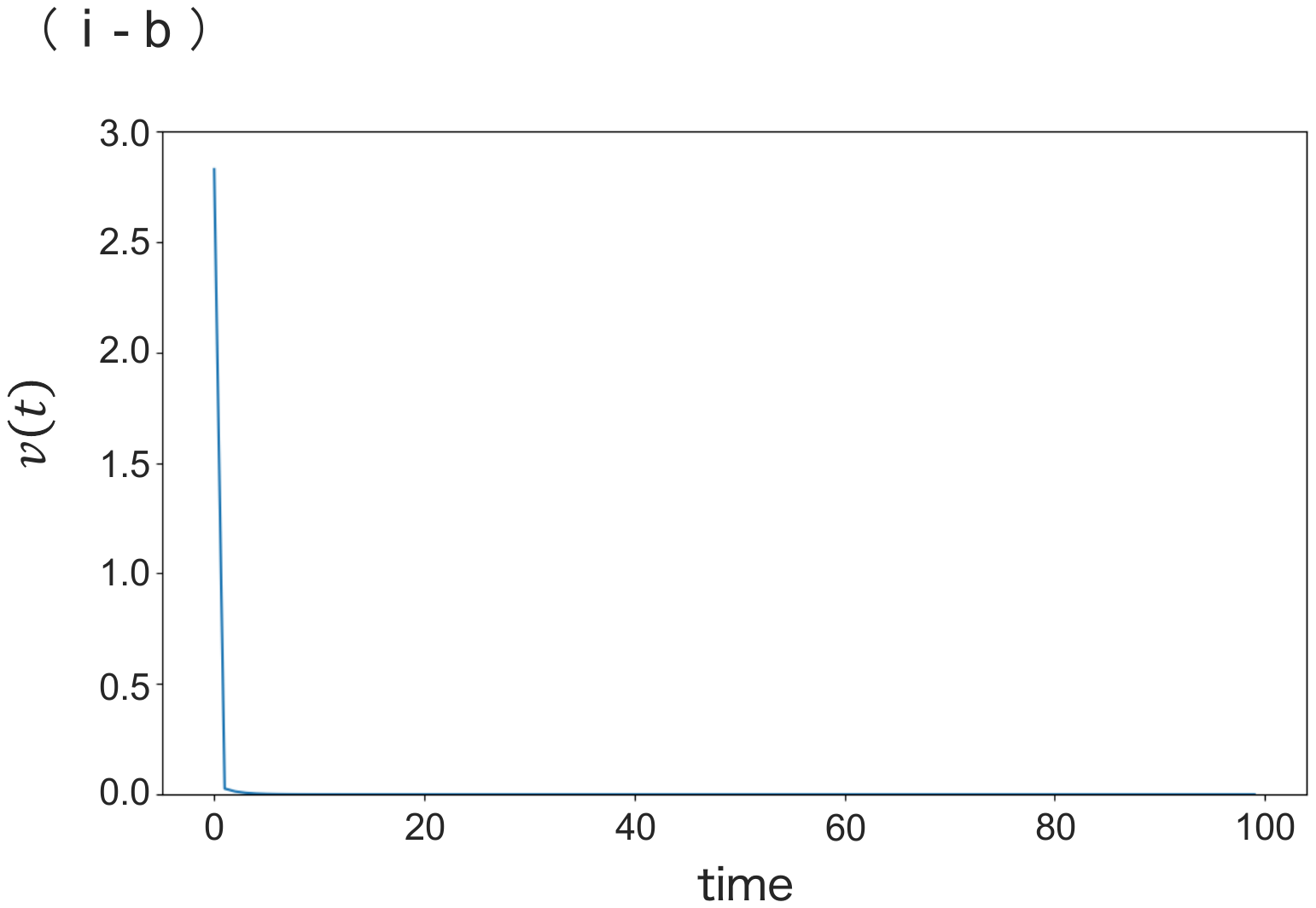}
      \end{minipage}
    \end{tabular}
\caption{
Comparison between the the simple case of 
stochastic viscous-motion model
\eqref{eq:simple_viscous}
and the dynamical system 
\eqref{eq:dynamics3}.
(i-a) for the stochastic model with $\widehat{\ell}=0.04$
and (ii-b) for the dynamical system with $\ell=0.01$.
}
\label{fig:st_dy1}
\end{figure}
\begin{figure}[ht]
    \begin{tabular}{cc}
      \begin{minipage}[ht]{0.4\hsize}
        \centering
        \includegraphics[keepaspectratio, scale=0.25]{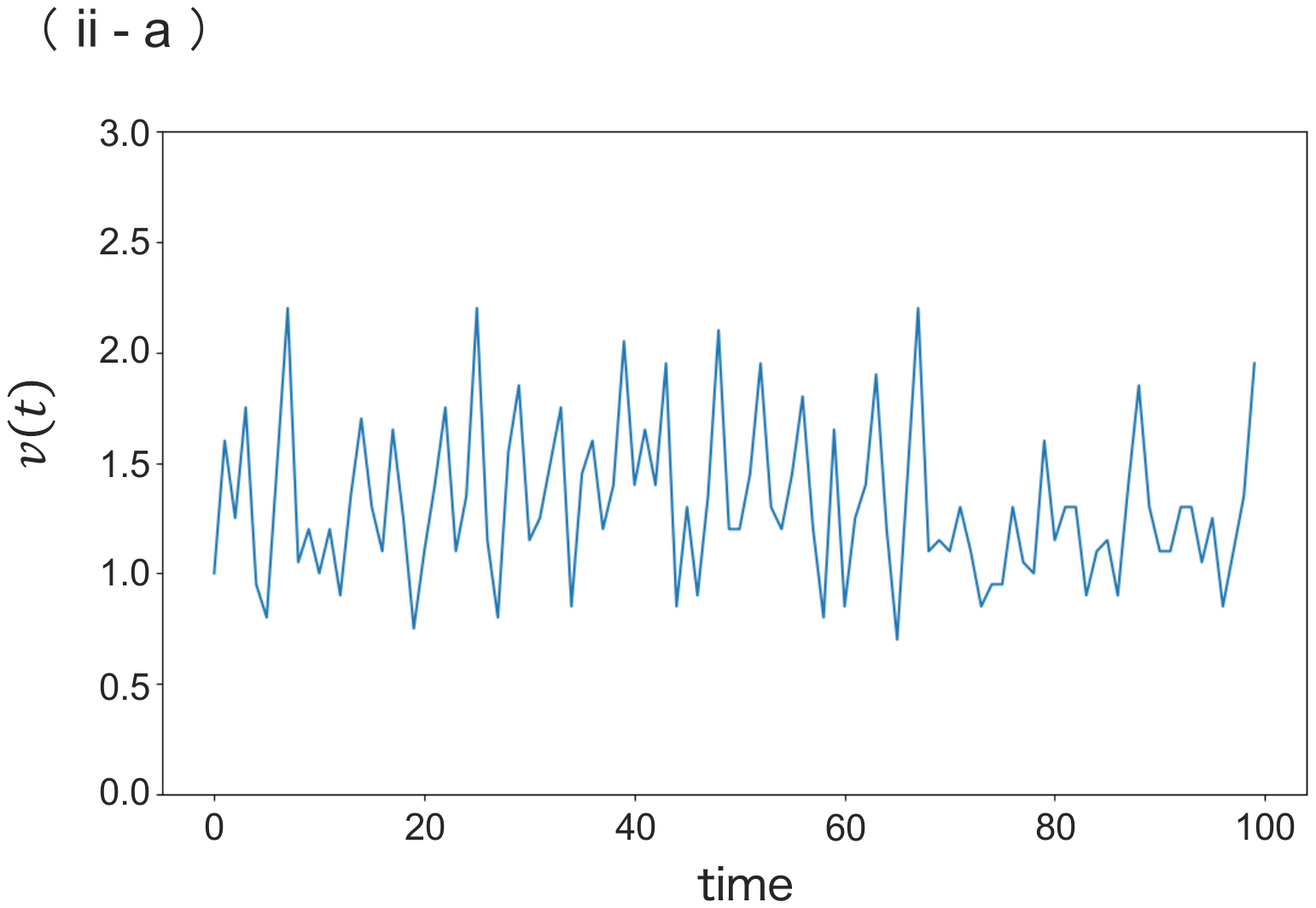}
      \end{minipage} &
      \begin{minipage}[ht]{0.4\hsize}
        \centering
         \includegraphics[keepaspectratio, scale=0.25]{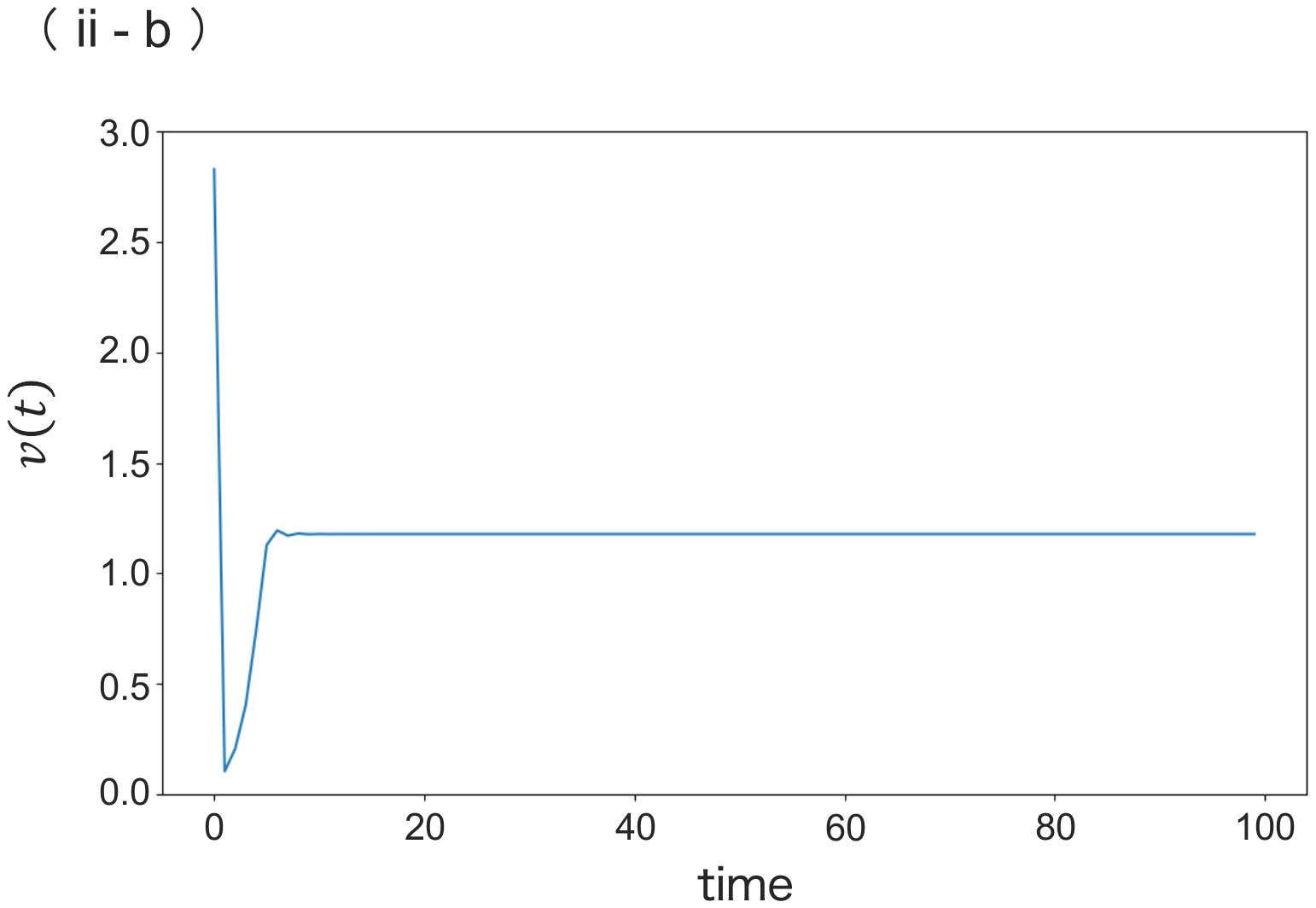}
      \end{minipage}
    \end{tabular}
\caption{
Comparison between the the simple case of 
stochastic viscous-motion model
\eqref{eq:simple_viscous}
and the dynamical system
\eqref{eq:dynamics3}.
(ii-a) for the stochastic model with $\widehat{\ell}=0.2$
and (ii-b) for the dynamical system with $\ell=0.1$.
}
\label{fig:st_dy2}
\end{figure}
\begin{figure}[ht]
    \begin{tabular}{cc}
      \begin{minipage}[ht]{0.4\hsize}
        \centering
        \includegraphics[keepaspectratio, scale=0.25]{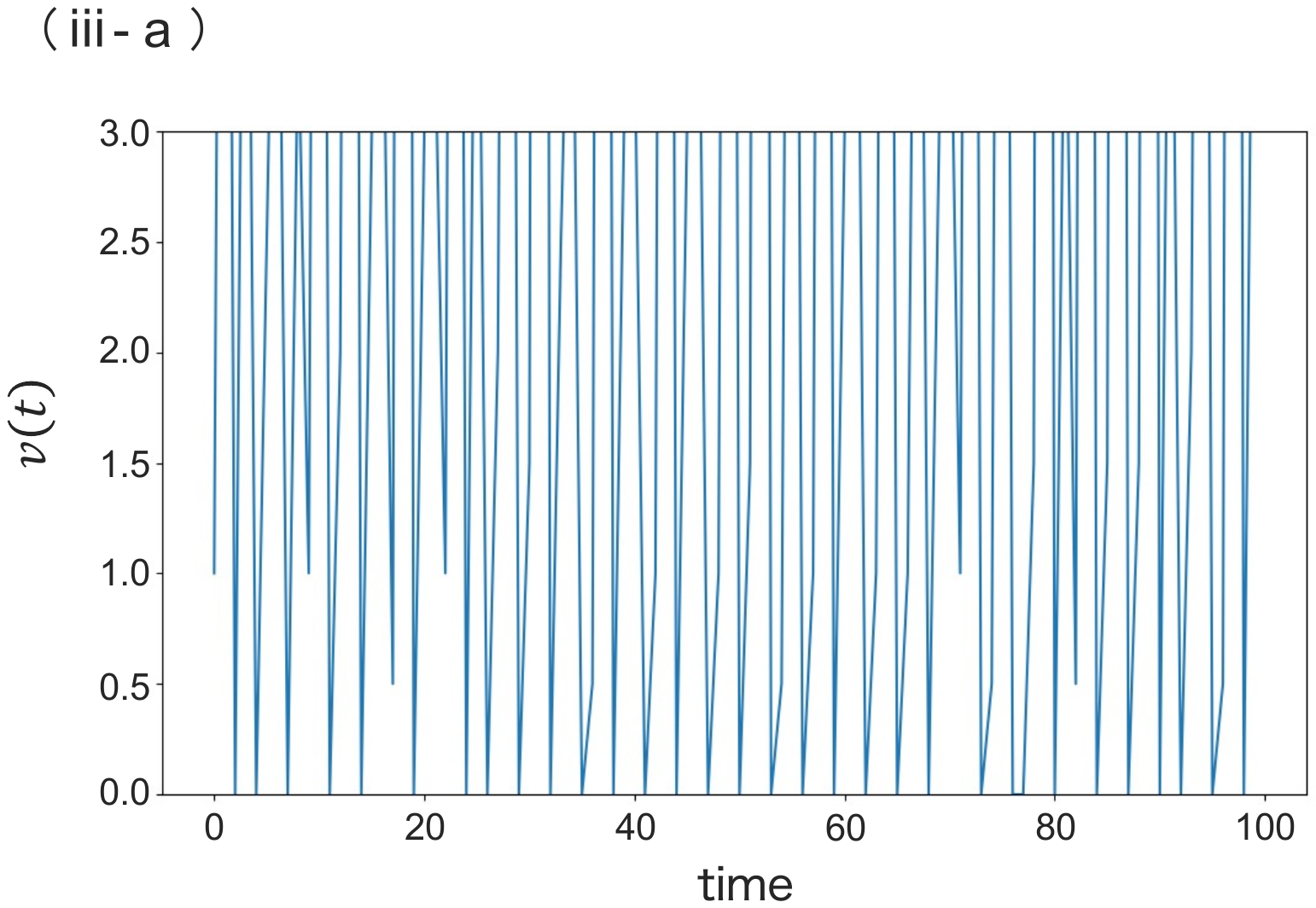}
      \end{minipage} &
      \begin{minipage}[ht]{0.4\hsize}
        \centering
         \includegraphics[keepaspectratio, scale=0.25]{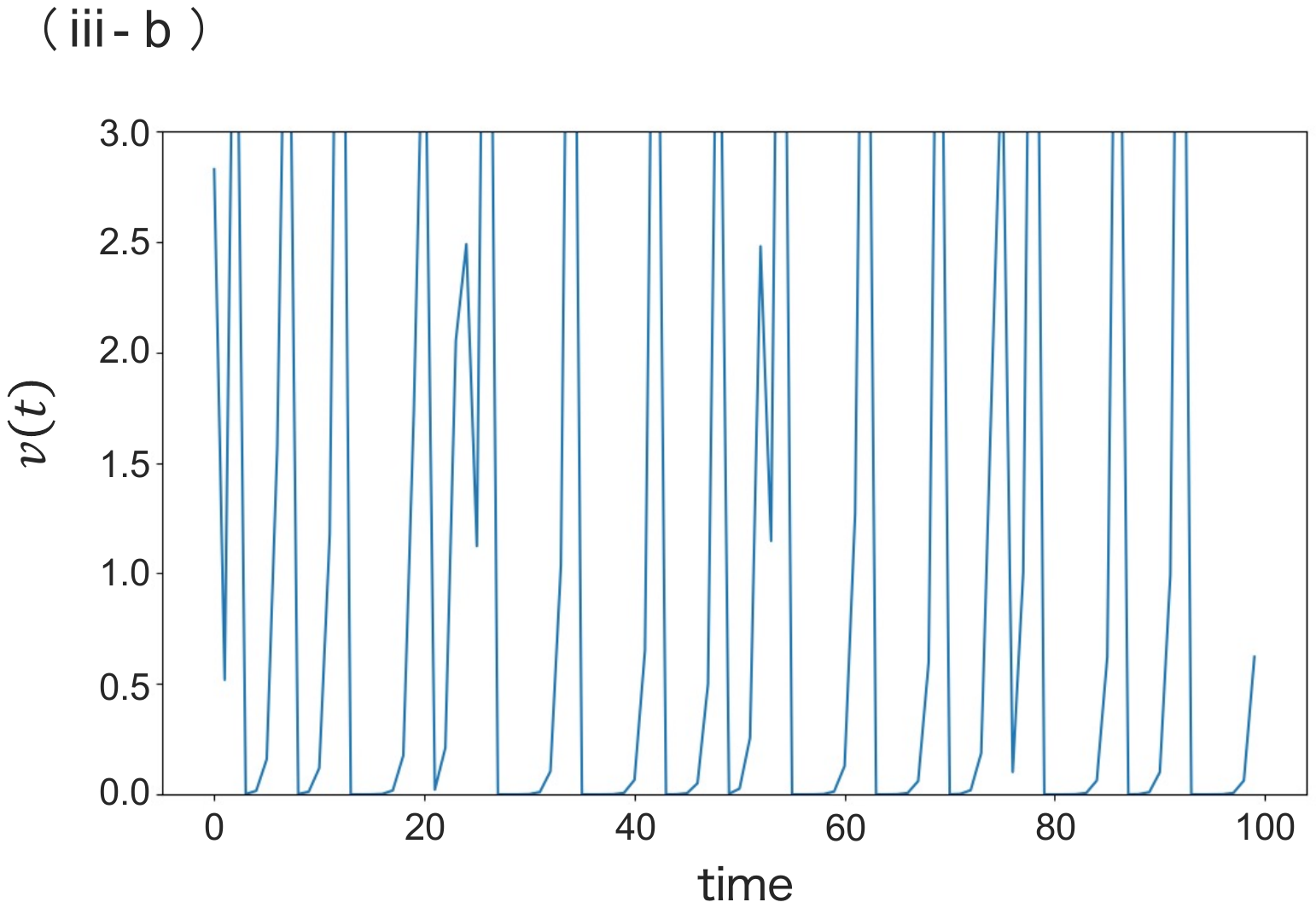}
      \end{minipage}
    \end{tabular}
\caption{Comparison between the the simple case of 
stochastic viscous-motion model
\eqref{eq:simple_viscous}
and the dynamical system
\eqref{eq:dynamics3}.
(iii-a) for the stochastic model with $\widehat{\ell}=2$
and (iii-b) for the dynamical system with $\ell=0.5$.
}
\label{fig:st_dy3}
\end{figure}

Here we compare the results of 
numerical simulation of 
the simple case of 
stochastic viscous-motion models
\eqref{eq:simple_viscous}
in Section \ref{sec:stochastic_viscous}
and the corresponding dynamical systems
\eqref{eq:dynamics3} in Section \ref{sec:dynamical}.
In the former, we define the parameter
\begin{equation}
\widehat{\ell}:= \frac{\beta L_x^2}{\mu}.
\label{eq:parameter1}
\end{equation}
Other parameters are set as mentioned in
Section \ref{sec:RW}; 
$N=50$, $m_1=m_2=5$, and $L_x=2$.
In the latter, we set
\begin{equation}
\alpha=N=50, \quad D=\frac{m_2}{2 m_1}=\frac{1}{2},
\label{eq:parameter2}
\end{equation}
and change the parameter $\ell$.
In this case, \eqref{eq:C} gives
\begin{equation}
C=\frac{\sqrt{2 \pi}}{50 \ell}
\fallingdotseq 5.0 \times 10^{-2} \times
\frac{1}{\ell}.
\label{eq:C2}
\end{equation}
In Figs.~\ref{fig:st_dy1}--\ref{fig:st_dy3} we show 
the time evolution of the velocity $v(t)$
for the stochastic viscous-motion models
starting from $v(0)=1$
in the left three figures
(in the time unit $T_1$), 
and
for the dynamical systems
starting from $v(0)=2 \sqrt{2}$.
in the right three figures.
The parameters are shown in the caption.
Notice that for the dynamical systems,
$\ell=0.01$ in Fig.~\ref{fig:st_dy1} (i-b) 
gives $C \fallingdotseq 5$ 
which satisfies the condition
of the case (i) in Proposition \ref{thm:stat}, 
$\ell=0.1$ in Fig.~\ref{fig:st_dy2} (ii-b) 
gives $C \fallingdotseq 0.5$
which satisfies the condition
of the case (ii) in Proposition \ref{thm:stat}, 
and
$\ell=0.5$ in Fig.~\ref{fig:st_dy3} (iii-b) 
gives $C \fallingdotseq 0.1$
which satisfies the condition
of the case (iii) in Proposition \ref{thm:stat}.
Figure~\ref{fig:st_dy1} (i-a) and (i-b) show that
the rest state $v(t)=0$ is stable,
and Fig~\ref{fig:st_dy2} (ii-a) and (ii-b) show that
the systems have the non-zero velocity states
$v(t) \not=0$ as the steady state.
On the other hand, Fig~\ref{fig:st_dy3} (iii-a) and (iii-b) show
oscillatory behavior of motions between
$v=0$ state and $v \not=0$ state.
In summary, the transitions between the three
different phases of motions are well-described
by the dynamical systems \eqref{eq:dynamics3},
and the stochastic viscous-motion models
can show fluctuations around the
solutions of the dynamical system.

\SSC
{Concluding Remarks and Future Problems}
\label{sec:future}

We list out the remarks and future problems.
\begin{description}
\item{(i)} \, 
We have analyzed the stochastic models
and the dynamical systems in the viscosity dominated
cases for one camphor disk.
The time evolution of the mean values of
velocity is well described by the dynamical system
for each values of parameters.
The fluctuations around the mean values 
simulated by the stochastic models
shall be studied systematically 
to clarify the parameter dependence of
the distribution of deviations from mean values.
Moreover, systems including 
two or more than two disks in one-dimension
should be analyzed in order to understand the
interactions among camphor disks
though the camphor molecules in water
which are simulated by random walks
in the present stochastic models.
The effect of acceleration and deceleration 
will be studied by the stochastic Newtonian-motion
models introduced in Section \ref{sec:RW}.
In Section \ref{sec:comparison}, we fixed parameters
different from $\widehat{\ell}=\beta L_x^2/\mu$.
More systematic study on the parameter dependence
should be studied.

\item{(ii)} \quad
As demonstrated in Section \ref{sec:simulation},
two- and three-dimensional systems including many 
camphor disks seem to be quite interesting.
The simulations of the proposed stochastic
Newtonian-motion models shall be done
and the results should be compared with
the experiments reported by the 
literature \cite{MINS19,NSN17,STNN15}.
In particular, setting of proper boundary conditions,
e.g., a circular boundary condition
with changing radius, 
will be important to make comparison between
the experimental results and the theoretical 
study using stochastic and dynamical systems.
\end{description}

In the present work, we have concentrated on the
motions of camphor disks interacting though
heterogeneous concentration-field
of camphor molecules.
Here the filed of camphor molecules is
generated and changed by the 
camphor disks and also
the time-evolution of the filed 
affects the motion of camphor disks.
Such feedback effects are common in
self-propelled systems.
In the present case, the field causes 
the repulsive interaction 
among disks \cite{MINS19,NSN17,STNN15},
but we will be able to apply the present 
models to the self-propelled particle
systems with attractive interactions.
We are considering the possibility to
apply our models to study
the collective motions of the eusocial insects,
in particular, of ants \cite{EMTKN24,MKN25}.
Another possibility is to apply the present
study to the traffic flow and jam problems,
where the road conditions and 
interaction between drivers shall be
expressed by the heterogeneous 
and time-dependent field around cars.



\end{document}